\documentclass[aps,twocolumn,pra,showpacs]{revtex4}
\usepackage{amsfonts}
\usepackage{mathrsfs}
\usepackage{graphicx}
\usepackage{amsmath,amssymb,amsthm,latexsym}
\usepackage{amsthm}
\usepackage{amsbsy}
\usepackage{epsfig}
\usepackage{graphicx}
\usepackage{float}
\usepackage{dcolumn}
\usepackage{amsmath}

\begin{document}

\author{Yong-Cheng Ou$^1$, C. Allen Bishop$^1$, and Mark S. Byrd$^{1, 2}$}

\affiliation{$^1$Physics Department and $^2$Computer Science
Department, Southern Illinois University, Carbondale, Illinois
62901-4401 }

\title{Sufficient conditions placed on initial system-environment states for positive maps }

\begin{abstract}
A system interacting with its environment will give rise to a 
quantum evolution. After tracing over the environment the net
evolution of the system can be described by a linear 
Hermitian map.  It has recently been shown that a necessary and
sufficient condition for this evolution to be completely positive is
for the initial state to have vanishing quantum discord.  
In this paper, we provide a sufficient condition
for the map to be positive with respect to the initial
system-environment correlation.  This could lead to ways in which to
identify positive but not completely positive maps.  Illustrative
examples and suggestive procedures are also provided.  
\end{abstract}


\maketitle


\section{Introduction}

Any quantum system will inevitably interact with its
environment in some way. Since the environment is generally not
available to us, it is the system alone which is typically observed or
measured.  As a result the system is open and does not evolve
in a unitary fashion. Dynamical maps were proposed to describe the
state of a system \cite{Sudarshan:61}.  The maps can be classified as being
either positive or non-positive, with the positive maps including
completely positive (CP) maps.  

Suppose the system $A$ interacts with an 
environment $E$. After an evolution determined by the standard
quantum-mechanical prescription, its density matrix at a given time
$t$ will reduce to 
\begin{eqnarray} \nonumber
  \rho_A(t) &=& \texttt{Tr}_E[\rho_{AE}(t)]\\ \nonumber
   &=& \texttt{Tr}_E[U_{AE}(t)\rho_{AE}(0)U_{AE}(t)^{\dag}] \\ \label{a1}
   & \equiv & \mathcal {S} [\rho_A(0)],
\end{eqnarray}
where $U_{AE}(t)$ is a unitary matrix determined by the joint
system-environment Hamiltonian, $\mathcal {S}$ is the induced map,
and $\rho_A(0)=\texttt{Tr}_E \rho_{AE}(0)$.  In recent years there has
been an extensive investigation regarding the conditions imposed on
the initial state of a composite system which lead to either positive
or CP maps \cite{Pechukas:94,Pechukas+Alicki:95,Cesar/etal:08,Jordan:04,Shabani/Lidar:09a}.

 It is well known that if the initial state $\rho_{AE}(0)$ is of a
 simple product form, 
 i.e., $\rho_{AE}(0)=\rho_A\otimes|0\rangle _E \langle 0|$, the
 resulting map $S$ is CP \cite{Nielsen/Chuang:book}. 
 Simply separable states are not the
 only ones whose evolution can be described by a CP map \cite{Cesar/etal:08}, the
 general class consists of those states with vanishing quantum discord
 (VQD) \cite {Oll/Zurek:01}. It has been shown that such a quantum dynamical
 process (\ref{a1}) always leads to a linear  
 Hermitian map $S$, and for arbitrary $U_{AE}(t)$ the initial state
 with VQD is not only sufficient  \cite{Cesar/etal:08} but also necessary for CP maps
 \cite{Shabani/Lidar:09a}.  
 Positive but not CP maps play an important role in
 detecting entanglement of quantum states \cite{Peres,Horodeckis}. Using matrix algebras, some positive maps were constructed
 in Ref.~\cite{Chrus/Kossakowski}.  However, with the exception of the CP maps, we know
 little about the condition(s) which must be imposed on an initial 
 state so that the subsequent evolution is a positive map for arbitrary
 $U_{AE}$.  

 In this paper, we will give a sufficient condition for the maps 
 $\mathcal{S}$ (\ref{a1}) to be positive with respect to the initial 
 composite state and conjecture this condition is necessary as 
 well. This result, together with that of
 \cite{Cesar/etal:08,Shabani/Lidar:09a} may provide an
 efficient way of finding some positive but not CP maps.


\section{Sufficient Conditions for Positivity}

 A separable quantum state $\rho _{AE}$ on $\mathcal {H}_A \otimes \mathcal
 {H}_E$ with $d \otimes f$ dimensions can be expressed as a convex
 combination of product states \cite{Werner:89}, i.e., in the form
 \begin{equation} \label{a2}
\rho_{AE}=\sum_i p_i\rho_A^{(i)} \otimes \rho_ E^{(i)},
\end{equation}
with nonnegative $p_i$ satisfying $\Sigma_i p_i =1$.  
The state (\ref{a2}) can be rewritten as
\begin{equation} \label{a3}
\rho_{AE}=\sum _{kl} \Gamma _{kl} |k\rangle\langle l|\otimes
\psi_{kl},
\end{equation}
where $\{|k\rangle \}_{k=1}^{d}$ represents an orthonormal basis for
the Hilbert space of system $A$, 
$\mathcal {H}_A$, and $\{\psi_{kl}\}_{k, l=1}^{d} : \mathcal
{H}_E\mapsto \mathcal {H}_E$ are normalized such that if 
$\texttt{Tr}[\psi_{kl}]\neq 0$ then $\texttt{Tr}[\psi_{kl}]= 1$. The
reduced density matrix of the system $A$ is
\begin{equation}
\rho_A=\sum_{(k,l)\in \mathcal {C}}\Gamma _{kl} |k\rangle\langle l|,
\end{equation}
where $\mathcal {C} \equiv \{(k,l)|\texttt{Tr}[\psi_{kl}]=1\}$.  The
special-linear (SL) class of states \cite{Shabani/Lidar:09a} is defined such that 
$\texttt{Tr}[\psi_{kl}]=1$ or $\psi_{kl}=0, \forall k, l$.  Furthermore,
for the SL class we have $\Gamma _{kl}\neq 0$ for $\psi_{kl}$
with $\texttt{Tr}[\psi_{kl}]=1$ or $\psi_{kl}\neq 0$, while $\Gamma
_{kl}= 0$ for $\psi_{kl}=0$.  

We denote the elements of component $\rho _A ^{(i)}$ in (\ref{a2})
by $\mathscr{E}_{kl}^{(i)}$, i.e., 
$\rho_A^{(i)}=\sum_{kl}{\mathscr{E}}_{kl}^{(i)}|k\rangle\langle l|$,
such that for the separable SL class the 
bath operator $\psi_{kl}$ can be written 
\begin {equation} \label{b2}
\psi _{kl}=\sum _{i}\frac{p_i\mathscr{E}_{kl}^{(i)}}{\Gamma
_{kl}}\rho _E ^{(i)},
\end {equation}
with $\Gamma_{kl}=\Sigma_{i}p_i \mathscr{E}_{kl}^{(i)}$ for $\Gamma
_{kl}\neq 0$. Rewriting the dynamical map (\ref{a1}) we have
\begin {equation}
\mathcal {S} [|k\rangle \langle l
|]=\texttt{Tr}_E[U_{AE}(t)(|k\rangle \langle l |\otimes
\psi_{kl})U_{AE}(t)^{\dag}],
\end{equation}
for the SL class. On the other hand, there is a shift term which is 
independent of $\rho_A$ for the non-SL class \cite {Shabani/Lidar:09a}. If
$\psi_{kl}=0$, $\mathcal {S} [|k\rangle \langle l |]=0$, i.e., the
corresponding basis $|k\rangle \langle l |$ has no contribution to 
the resulting state. In what follows we will assume that the system
and bath are initially in a separable SL class state.

In order to find the condition under which the map $\mathcal
{S}$ is positive, we need to apply $\mathcal {S}$ to an arbitrary
$d\times d$ density matrix 
\begin{equation}
\rho_A'=\sum_{kl}\Gamma _{kl}' |k\rangle\langle l|,
\end{equation} 
to see whether the resulting matrix
\begin{eqnarray}
  \mathcal {S}[\rho'_A] &=& \sum _{kl}\texttt{ Tr}_E[U_{AE}(t)(\Gamma
_{kl}' |k\rangle \langle l |\otimes \psi_{kl})U_{AE}(t)^{\dag}],\label {b3}
\end{eqnarray}
is positive or not. Let us define a set of matrices as
\begin{equation} \label {e3}
\varrho_A ^{(i)}=\sum_{kl} \frac{\Gamma
_{kl}'\mathscr{E}_{kl}^{(i)}}{\Gamma _{kl}}|k\rangle\langle l|\equiv
\sum_{kl}\Gamma _{kl}'\Gamma^{(i)}_{kl}|k\rangle\langle l|,
\end{equation}
with $ \Gamma^{(i)}_{kl}=\mathscr{E}_{kl}^{(i)}/\Gamma _{kl}$.  
Using (\ref{b2}) and (\ref{e3}) we can reexpress (\ref{b3}) as 
\begin{eqnarray} 
  \mathcal {S}[\rho'_A] &=& \sum _{ikl} p_i
\texttt{Tr}_E[U_{AE}(t)(\Gamma _{kl}'\Gamma_{kl} ^{(i)} |k\rangle
\langle l |\otimes \rho _E ^{(i)})U_{AE}(t)^{\dag}] \nonumber \\
 &=& \sum _{i}p_i \texttt{Tr}_E[U_{AE}(t)(\varrho_A ^{(i)}\otimes 
      \rho_E^{(i)})U_{AE}(t)^{\dag}]. \label{ui}
\end{eqnarray}

From (\ref{ui}) it is apparent that if the matrix $\varrho_A^{(i)}\geq 0,
\forall i$, $ \mathcal {S}[\rho'_A]$ will be positive, the sum of positive
density matrices is indeed positive. Since $\rho'_A$ represents an
arbitrary density matrix, having $\varrho_A ^{(i)}\geq 0$ for $\forall
i$ implies the mapping $ \mathcal {S}$ will be positive as well.

Before proceeding, let us state the following Lemma.

Lemma 1: For two positive matrices defined by $ \rho_1=\sum
_{ij}\phi_{ij} |i\rangle\langle j|$ and $ \rho_2=\sum
_{ij}\varphi_{ij} |i\rangle\langle j| $, there exists an
unnormalized matrix $\rho$ such that
\begin{equation} \label{h12}
\rho \equiv \sum _{ij} \phi_{ij}\varphi _{ij} |i\rangle\langle j|
\geq 0.
\end{equation}
Proof: This proof is straightforward. Since $\rho_1\otimes \rho_2$
is nonnegative, its principal submatrix is also nonnegative. It is
clear that the matrix $\rho$ is a principal submatrix of
$\rho_1\otimes \rho_2$. Thus (\ref{h12}) follows.

Using this Lemma, 
the comparison between (\ref{e3}) and (\ref{h12}) now provides us with
a condition for the $\varrho_A^{(i)}$ to be positive.  If the
re-scaled matrices 
\begin{equation} \label {r1}
\varrho_{R}^{(i)}\equiv \sum_{kl} \Gamma^{(i)}_{kl}|k\rangle\langle
l|, \forall i
\end{equation}
are all nonnegative, then $\varrho_A ^{(i)}\geq 0, \forall i$, and
thus $ \mathcal {S}[\rho'_A] \geq 0$. Note that the equation above is
only dependent on the initial state. Therefore, we 
can draw the conclusion that for an arbitrary $U_{AE}$, the map
$\mathcal {S}$ defined by (\ref{a1}) is positive if the initial
system-bath state $\rho_{AE}$ is in the separable SL class and all of its
re-scaled matrices (\ref{r1}) are nonnegative.  

It is not difficult to verify that the conclusion above may be
reformulated in the following way, which constitutes the main result of
this paper.

\textbf{Theorem}: For an arbitrary $U_{AE}$, the map $\mathcal{S}$
defined by (\ref{a1}) is positive if the initial system-bath SL state
$\rho_{AE}$ in $d \otimes f $ dimensions is of the unentangled form
(\ref{a2}) and the component $\rho_A ^{(i)}$ can be written as
\begin {equation}
\rho _A ^{(i)}=\Pi _{i}^ {(d_i)}\rho _A ^{(i)}\Pi _{i}^ {(d_i)},
\end {equation}
where $\{\Pi _{i}^ {(d_i)}\}$ are $d_i$-dimensional projectors onto
$\rho _A^ {(i)}$ and $\sum_i \Pi _i ^{(d_i)} = I_d$ with $\Sigma _i
{d_i}=d$.


\subsection{Example}

For an intuitive picture of the theorem, an illustration is given by
 the following example. Consider the separable initial $4\otimes f$
state
\begin {equation} \label{p1}
\rho_{AE}=p_1\rho_A^{(1)}\otimes \rho_E^{(1)}+p_2\rho_A^{(2)}\otimes
\rho_E^{(2)},
\end{equation}
where $\rho_A^{(i)}$ is of the form in the computational basis
$\{|0\rangle |1\rangle\ |2\rangle\ |3\rangle\}$
\begin {equation}
\rho_A^{(1)}= \left(
               \begin{array}{cccc}
               a& b & 0 &0\\
                c & d &0 & 0 \\
                0& 0 & 0 & 0 \\
                0& 0 & 0 & 0 \\
               \end{array}\right)
            \equiv \sum _ {k,l =1}^{4}\mathscr{E}_{kl}^{(1)}|k\rangle\langle
            l|,
\end{equation} and
\begin {equation} \nonumber
\rho_A^{(2)}= \left(
               \begin{array}{cccc}
               0& 0 & 0 &0\\
                0 & 0 &0 & 0 \\
                0& 0 & e & f \\
                0& 0 & g & h \\
               \end{array}\right)
            \equiv \sum _ {k,l =1}^{4}\mathscr{E}_{kl}^{(2)}|k\rangle\langle l|.
\end{equation}
It is easy to check that $\rho_{AE}$ is a SL state and the reduced
state of subsystem $A$ is

\begin {equation} \nonumber
\rho_A=\left(
               \begin{array}{cccc}
                 p_1a & p_1 b & 0 & 0\\
                 p_1c & p_1 d & 0 & 0\\
                 0 & 0& p_2e & p_2 f\\
                 0 & 0 & p_2 g & p_2 h \\
               \end{array}
             \right)
    \equiv \sum _ {k,l =1}^{4} \Gamma _{kl} |k \rangle\langle l|.
\end{equation}
According to (\ref{e3}) and (\ref{r1}) we can have the re-scaled
matrices as follows
\begin {equation} \nonumber
\varrho_{R}^{(1)}=\sum _ {k,l =1}^{4} \Gamma _{kl}^{(1)} |k
\rangle\langle l| \\\nonumber
     =\left(
               \begin{array}{cccc}
                 \frac{1} {p_1} & \frac{1} {p_1}& 0 & 0\\
                 \frac{1} {p_1} & \frac{1} {p_1} & 0 & 0\\
                 0 & 0 & 0 & 0\\
                  0 & 0 & 0 & 0\\
               \end{array}
             \right),
\end{equation}
and
\begin {equation} \nonumber
\varrho_{R}^{(2)}=\sum _ {k,l =1}^{4} \Gamma _{kl}^{(2)} |k
\rangle\langle l| \\\nonumber
     = \left(
               \begin{array}{cccc}
                 0 & 0 & 0 & 0\\
                 0 & 0& 0 & 0\\
                 0 & 0 & \frac{1}{p_2} & \frac{1}{p_2}\\
                 0 & 0 & \frac{1}{p_2} & \frac{1}{p_2}\\
               \end{array}
             \right).
\end{equation}
The two re-scaled matrices above are obviously nonnegative such that
the map (\ref{a1}) resulting from the initial state (\ref{p1}) is
positive for arbitrary $U_{AE}$.


\subsection{Discussion}

Note that if all $d_i=1$ in the theorem, then the state has a VQD and
the map must be a CP map \cite{Shabani/Lidar:09a}.  (For example if the 
$\rho_A^{(i)}$ in Eq.~(\ref{p1}) in the example are 1-D projectors.) 
However, if the state Eq.~(\ref{a2}) does not have VQD it may be
possible to find $U_{AE}$ such that the map is not 
CP.  Therefore, the theorem could provide us a way to search
 for positive but not CP maps by varying $U_{AE}$.  Since the set of
 CP maps is a subset of the positive maps, the restriction to the
 initial states with respect to positive maps is relaxed, compared to
 one for the CP maps.  

It could be the case that the sufficient condition given in the theorem for
positive maps is necessary as well. However, it is predicted that 
one shall encounter more challenge giving a complete proof and thus
it deserves further investigation. In the following, we provide an 
explicit example showing that an initial entangled composite state 
can lead to a non-positive map by using the analysis of
\cite{Cesar/etal:08,Shabani/Lidar:09a}.  

Consider the initial state in the entangled form
\begin{equation} \label{p}
\rho_{AE}=\frac{1}{\sqrt{2}}(|00\rangle+|11\rangle).
\end{equation}
Since the state does not belong to the SL class, the resulting map
is not necessarily positive. Indeed, the following computation 
verifies this fact.

As shown by \cite{Shabani/Lidar:09b}, the map $\mathcal {S}(t)$ induced from
(\ref{p}) assumes an affine form  
\begin{equation} \label{l1}
\mathcal {S}{(t)}[\rho_E(0)]=\mathcal {S}_{SL}{(t)}[\rho_E(0)]+\mathcal {S}_{NSL}(t),
\end{equation}
where $\mathcal {S}_{SL}(t)$ depends on $\rho_E(0)$ while $\mathcal
{S}_{NSL}(t)$ does not. (For more details on $\mathcal {S}_{SL}(t)$,
see \cite{Shabani/Lidar:09b}.)  Next, we try to apply the map $\mathcal {S}(t)$ to a
particular pure-state density matrix  
\begin{equation}
\rho'= \left( \begin{array}{cc}
                 1 & 0  \\
                 0 & 0  \\
               \end{array}\right).
\end{equation}
Clearly, this matrix is positive semi-definite. From (\ref{b3}) 
\begin{equation}
\mathcal {S}_{SL}{(t)} [\rho']=\frac{1}{2}\rho'.
\end{equation}
For the state (\ref{p}), the shift term $\mathcal {S}_{NSL}(t)$ in
(\ref{l1}) has the form \cite{Shabani/Lidar:09a} 
\begin{equation}
\mathcal {S}_{NSL}(t)=\sum _{kl \in \{01, 10\} } \Gamma _{kl} \texttt {Tr}_E[U_{AE}(t)|k\rangle\langle l|\otimes \psi _{kl} U_{AE}(t)^{\dag}],
\end{equation}
where $\Gamma _{01}=\Gamma_{01}=\frac{1}{2}$, and
\begin{equation}
 \psi _{01}= \left( \begin{array}{cc}
                 0 & 1  \\
                 0 & 0  \\
               \end{array}\right), \psi _{10}= \left( \begin{array}{cc}
                 0 & 0  \\
                 1 & 0  \\
               \end{array}\right).
\end{equation}
Now we choose $U_{AE}(t)$ as a CNOT gate, i.e.,
\begin{equation}
U_{AE}(t)=|00\rangle\langle 00|+|01\rangle\langle 01|+|10\rangle\langle 11|+|11\rangle\langle 10|,
\end{equation}
such that we obtain resulting matrix
\begin {eqnarray} \nonumber
\mathcal {S}(t)[\rho']& = & \frac {1}{2}\rho ' + \mathcal
{S}_{NSL}(t)
\\\nonumber
     & = & \frac{1}{2} \left(
               \begin{array}{ccc}
                 1 & 1 \\
                 1 & 0  \\
               \end{array}
             \right),
\end{eqnarray}
which is negative. Therefore, we conclude that the map of the
state (\ref{p}) is not always positive for all $U_{AE}$.


\section{Conclusion}

In conclusion, we have obtained a sufficient condition
for positive maps with respect to a given initial system-environment
state. The positive but not CP maps are important for identifying
the entanglement of quantum states, and our result may
provide an efficient way of constructing such maps. This can be
performed as follows. First, we choose the initial state satisfying
the theorem above. Second, we exclude the states with the VQD.
Finally, we try a variety of unitary transformations until the
desired maps are found. Here, we need the method presented in \cite
{Jordan:04} to determine if the map is CP or not.  We leave as an important
open problem whether the condition is necessary as well, or which
condition(s) are necessary.  This would provide an improved method 
for searching for positive but not CP maps.


\end{document}